# Sink Mobility To Ensure Coverage in Multi-Partitioned Wireless Sensor Network

Hasina Rahman
*Computer Science and Engineering*
*Aliah University*
*Kolkata, India*
*Email: hasinarahman30@gmail.com*

Zeenat Rehena
*Computer Science and Engineering*
*Aliah University*
*Kolkata, India*
*Email: zeenatrehena@yahoo.co.in*

Nandini Mukherjee
*Computer Science and Engineering*
*Jadavpur University*
*Kolkata, India*
*Email:nmukherjee@jdvu.cse.ac.in*

***Abstract*—Recent works on WSNs show that use of mobile sink can prolong network lifetime. This paper demonstrates the advantages of the mobile sink in WSNs for increasing their lifetime than static sink. A novel sink mobility with coverage algorithm has been proposed here. During the movement of the sinks in the network, they sojourn at different points in the network and collect data over there. The direction of the movement of the sinks is determined by the algorithm and sinks are mobile inside the network. This algorithm can also be used for multiple sinks partitioned network. Simulations has been carried out for both mobile and static sinks to determine the average energy consumption of the network. At the same time, it also determines the network lifetime in terms of number of rounds neighbor nodes of sink is alive and first node die of the network. Our experiments show that mobile sinks outperform static sink in all scenarios. Furthermore, the proposed algorithm results in a good balancing of energy depletion among the sensor nodes.**

## 1. Introduction

The Wireless sensor network (WSN) is built of wireless sensor nodes of few hundreds to thousands. These sensor nodes are capable of sensing the physical quantity from surroundings. They can also process and store the acquired data and transfer them through wireless communication link to a sink. WSNs are dedicated for large range of applications, from monitoring (e.g. pollution prevention, precision agriculture, structures and buildings health) to event detection (e.g. intrusions, fire/flood emergencies), target tracking (e.g. surveillance), and recovery, control of industrial processes.

In the traditional WSN architectures, the networks were mostly assumed to be static. Most of the research papers focused on completely static scenarios. But, because of the intrinsic property of sensor nodes (small size, low weight, battery supply and wireless communication capability) sensor nodes are easily deployed on mobile entities, thus creating the potential for Mobile Wireless Sensor Networks (MWSN). Based on this observation, a large number of application scenarios have been designed. For instance, in the health-care application, patient monitoring is extremely important. Heart rate, blood pressure, body temperature etc. are common clinical parameters which must be constantly monitored. Use of portable systems based on WSNs makes it possible to constantly monitor these and other vital signs without restraining the patients, thus providing a more efficient and user-friendly medical service.

Further more, as opposed to static WSNs, mobile networks [1] are useful in several ways. Network connectivity becomes relaxed even in terms of nodes deployment since mobile elements can cope with isolated regions, and hence a sparse WSN becomes a feasible option. Thus mobile nodes cut down the necessity for dense WSN. Since static WSNs are dense and communication paradigm is often multi-hop, reliability is compromised by interface and collisions. In addition, the message loss increases with the number of hops, which may be rather high compared to mobile environment. On the other hand, in mobile WSN mobile elements visit nodes and collect data directly through single-hop transmissions. This not only reduces contention and collisions, but also reduces message loss. Moreover, in static WSN premature energy depletion of the nodes closer to the sink takes place due to overloading. This issue is known as funnelling effect [2], since the neighbors of the sink represent the bottleneck of traffic. Mobile elements in MWSN can reduce this funnelling effect, as they can visit different regions in the network and spread the energy consumption more uniformly. Again the cost of the network is reduced since fewer nodes can be deployed in a mobile WSN. Although mobility features of the node might be expensive, in many cases it is possible to exploit mobile elements such as trams, buses, shuttles or cars which are already present in the sensing area and attach sensors to them thus making the sensors mobile, as well as cutting down the cost. The main reason for which mobility is introduced in WSNs is to reduce the number of hops required to deliver data from sensor nodes to the base station. Thus, reducing the delay and prolonging the network lifetime by reducing the amount of energy required for data transmission.

In this paper we propose a sink mobility algorithm which finds the sojourn points of sink while moving inside the network. At the sojourn point the sink halts and collects data over there. This algorithm also ensures coverage of the network during mobility. To implement this algorithm, we assume that the network is already partitioned into four sub-networks and each sub-network has a sink for data communication. This is already done in our previous work [3]. Also MCDR [4] routing protocol is used to compare performances of mobile sink over static sink.

The rest of this paper is organized as follows. Section 2 presents some related works on sink mobility. Section 3 describes the proposed scheme and details the algorithm of sink mobility. Performance evaluation and results are demonstrated in Section 4. Finally, we conclude the paper in Section 5.

## 2. Related Work

Though several research works have dealt with the sink mobility in wireless sensor networks, the energy consumption of the nodes is the limiting factor for maximizing the network lifetime, since considerable energy is consumed not only for transmission of data but also for sensing, processing and hardware operations when in standby mode. Thus we have dealt with mobile sinks in a Wireless Sensor Network in this paper. In [5], authors introduced randomly moving mobile agents, called Data MULEs (Mobile Ubiquitous LAN Extension), which are used to collect data in sparsely populated sensor networks. On the other hand, Tong et al. have presented a similar solution in [6] for dense networks. They proposed SENMA architecture to exploit node redundancies by introducing mobile agents. A mobile agent is randomly flies above the sensor field and gathers data from sensors that are triggered based on the estimated fading state of their communication with the agent.

A predictable movement of the sink is also proposed in several papers. In [7], the sink moves along a predefined path, and takes data from the sensors when it comes closer to them. In [8] the strategy for sink mobility is different all together. The authors use adaptive sink mobility method to decrease energy consumption by. Sink is moved towards an optimal position and the approach is event-driven here. Multiple mobile sinks are considered in [9] for collecting data. Mobile sinks change their location when the nearby sensors energy becomes low. In this way the sensors located near sinks change over time. To decide a new location, a sink searches for zones which have richer sensor energy. Similar type of is shown in [10]. This work explores the idea of exploiting the mobility of data collection points (sinks) for the purpose of increasing the lifetime of a wireless sensor network. The authors give a linear programming formulation for the joint problems of determining the movement of the sink and the sojourn time at different points in the network. In [11], the authors consider a WSN with a mobile base station which repeatedly relocates its position to change the bottleneck nodes closer to the base-station. Multi-hop communication being used and the periphery of the covered circular region is used as a predefined path.

Similar approach is presented in [12], where authors consider a mobile base station which moves along a predetermined path. The use of controllably mobile elements to reduce the communication energy consumption.

Our work differs significantly from all the above mentioned previous works. We propose a novel sink mobility algorithm for ensuring coverage of the network. The propose work considers a multiple mobile sinks in partitioned network. In addition, compared to [9], when a sink moves, our algorithm searches for zones of sensors with high density to ensure coverage of the network. As a result, communication distance between farthest node and sink is reduced and movement of the sink depends on the topological changes of the network.

## 3. Proposed Scheme

In this section we explain our proposed algorithm for mobile sink in the WSN. The aim of this algorithm, Sink Mobility with Coverage (SiMoCo), is to move the sink along a certain path so as to cover all the sensor nodes while moving and thus collecting the sensed data, processing and analyzing them.

This algorithm is also designed for multiple sinks partitioned network [3]. The initial position of the sinks obtained from the CNP algorithm [13] which is discussed briefly in the next sub section. These positions are taken as the initial sojourn point of the sink. Then using the proposed algorithm, next subsequent sojourn points are calculated.

### 3.1. Centroid of the Nodes in a Partition (CNP) Algorithm

According to the CNP algorithm [13] sink node has location information. It is assumed that each partition has a single sink node and it has all location information of all nodes of a particular partition to which it belongs.

In CNP, the sink is initially placed at the centroid of all sensor nodes in a partition. Next the number of 1-hop neighbors of every sink is calculated. Then a new location of the sink is found by calculating the centroid of 1-hop neighbors for each partition. Again the number of 1-hop neighbors for new sink location is calculated. If the number of 1-hop neighbors of the new location is greater than the number of 1-hop neighbors of the old location, then new location becomes the sink location and again the previous steps are repeated until we finalize the sink location. That is when the number of 1-hop neighbors of the new location is found to be less than the number of 1-hop neighbors of the old location and the old location is taken as the final location of the sink.

Now the locations which are calculated using CNP is the initial sink locations according to the proposed algorithm in this paper. These initial positions of sinks in each partition are shown in Figure 1.

### 3.2. Sink Mobility with Coverage: SiMoCo

This section elaborately describes the SiMoCo algorithm. Initially the sinks are placed at a certain positions, $I$, which are obtained using CNP [13]. Next the nodes, $N_f$, i.e. the nodes which are not one-hop neighbors of the sinks are found and are placed within a buffer. The buffer also contains the location information of these nodes i.e. $L_{N_f}$. Again, the buffer is sorted in descending order according to the distance of the nodes from the initial position, $I$, of sink node. Next the farthest node is chosen from the buffer and the sink node is moved towards the node so that the farthest node comes within the range of the sink node to ensure coverage of the network using the following equations.

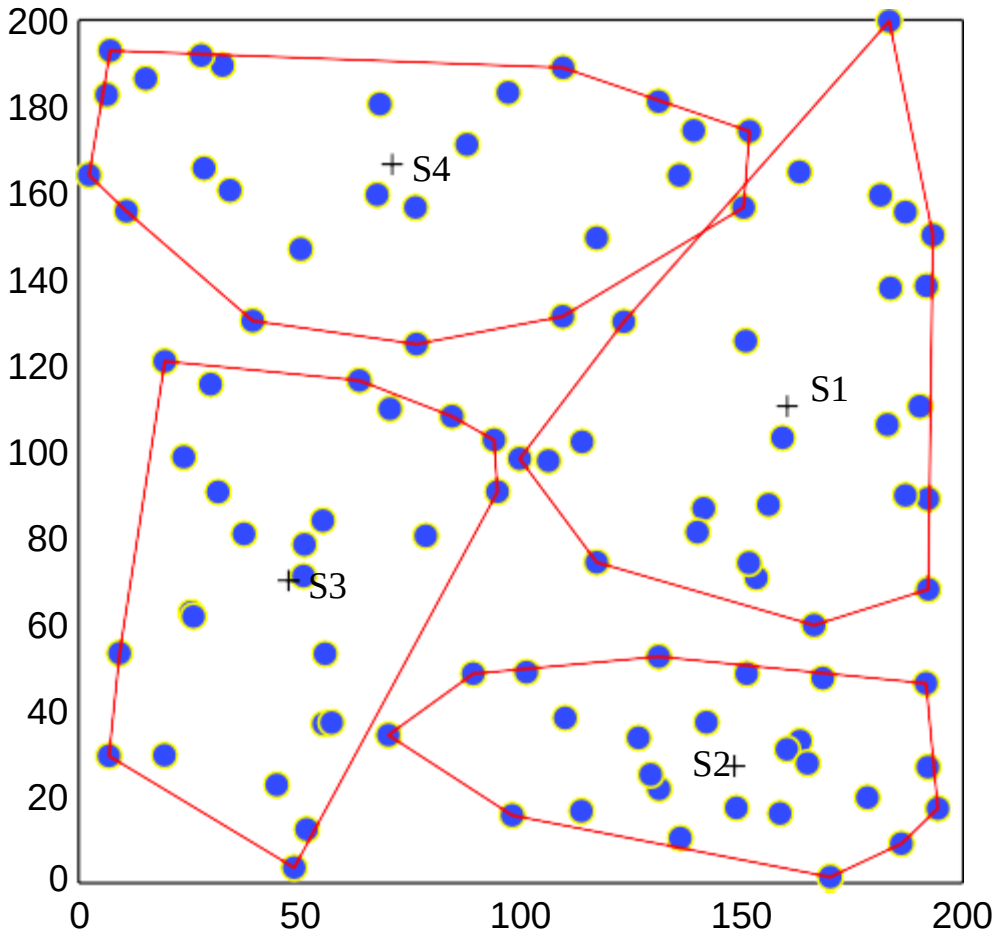

Figure 1. Initial Sink positions using CNP

$$P_k_x = ((move * X_P) + (range * X_{N_f j}))/D_{fs} \quad (1)$$

$$P_k_y = ((move * Y_P) + (range * Y_{N_f j}))/D_{fs} \quad (2)$$

Where $move$ is used to create a path between sink and the farthest node and $range$ is the communication distance. The $move$ is defined as $move = D_{fs}\ range$. Here $D_{fs}$ is the distance between sink and all the nodes other than one-hop neighbors of sink in each partition. $(X_P, Y_P)$ is the initial sink position and $(X_{N_f j}, Y_{N_f j})$ is $X$ and $Y$ coordinates of farthest node $j$ in buffer respectively.

After that, sink is moved to new location and again the one-hop neighbors are calculated. If the new one-hop neighbors are in buffer then these nodes are removed from the buffer. Next, the sink is again moved toward next farthest node in buffer and the previous steps are repeated until the buffer is empty. If the buffer is empty, the sink is moved to its initial location according to CNP algorithm. Table 1 describes the notations used in the algorithm and Algorithm 1 represents pseudo code of the SiMoCo algorithm. After using the SiMoCo in each partition, the sink is moved along the path which is shown in Figure 2.

## 4. Performance Evaluation

Performance of the sink mobility (SiMoCo) algorithm has been evaluated in MATLAB environment and compared with static sink positions in multiple sink environment. A square wireless sensor network is considered and network

TABLE 1. NOTATIONS USED IN THE ALGORITHM

| Notations | Descriptions |
|---|---|
| $N = \{N_1, N_2, ...N_n\}$ | n number of nodes |
| $X_i$<br>$Y_i$ | X coordinates of node $N_i$<br>Y coordinates of node $N_i$ |
| $N_f \in N$ | Contains farthest nodes that are not 1-hop neighbor nodes of sink |
| $D_{sn}$ | Distance between sink and all nodes |
| $D_{fs}$ | Distance between sink and all the nodes that are not 1-hop neighbors of sink |
| $P = \{P_1, P_2, ...P_k\}$ | k number of sinks |
| $B = \{D_{fs}, N_f, L_{N_f}\}$ | Buffer contains $D_{fs}$, $N_f$, and location info of nodes |
| $L_{N_f} = (X_{N_f j}, Y_{N_f j})$ | X coordinates of node $j$ in Buffer<br>Y coordinates of node $j$ in Buffer |

**Algorithm 1** Sink Mobility with Network Coverage

```
for each Partition k do
  Initialize sink location I = (X_P , Y_P )
  while ( B   NULL ) do
    Create a line between (X_P , Y_P ) & the 1st node of N_f ;
    move = D_fs − range;
    P_k_x = ((move * X_P ) + (range * X_Nfj ))/D_fs; P_k_
    y = ((move   Y_P*) + (range   Y_Nfj ))/D_fs;
    Calculate 1 hop neighbor;
    if N ∈ B then
      Remove it from B;
    end if
    X_P = P_k_x;
    Y_P = P_k_y;
  end while
  while ( B == NULL ) do
    move sink at initial location I;
  end while
end for
```

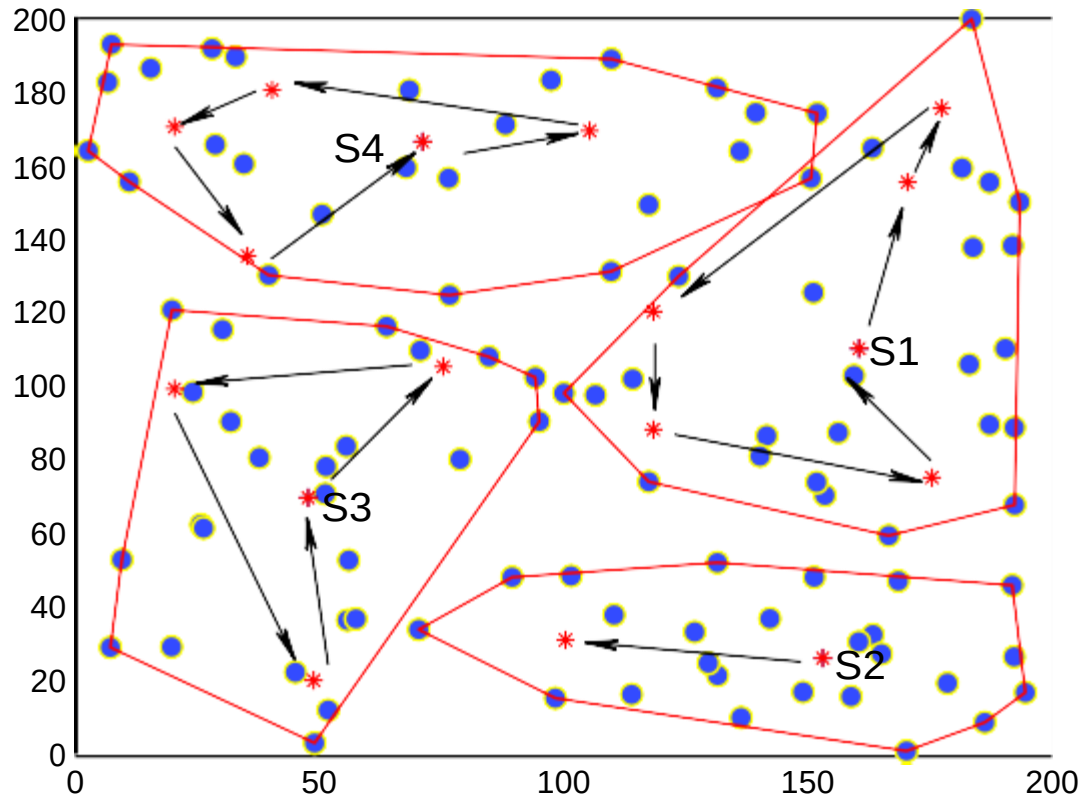

Figure 2. Sink movement after using the SiMoCo Algorithm

is partitioned into four sub-networks. Here 100 nodes are randomly deployed in a 200m X 200m square area assuming all nodes have same capabilities. In some experiments we use six different sensor networks ranging from 50 nodes to 300 nodes. The 50 node field is generated by randomly placing the nodes in a 200 m x 200 m square area. Other sizes are generated by scaling the square and keeping the communication range constant in order to keep average density of sensor nodes [6] constant. Each sink is placed in each partition. The communication range of the sensor nodes are assumed 45m and initially each node has same level of energy i.e. 0.5 joule. To calculate the transmission energy First Order Radio Model [14] is considered here. In both environment MCDR routing protocol is used to measure the different performance metrics.

### 4.1. Performance Metrics

The following metrics are used.

1) **Average Energy Consumption:** It is defined as the average energy is needed to transmit a packet from source to sink.
2) **Rounds taken to Neighbors Die:** It is calculated as the number of rounds is needed for all the one-hop neighbor nodes of sink to die.
3) **Rounds taken to First Node Die:** It is calculated as the number of rounds is needed for first node in the network to die.
These two metrics are used to demonstrate the lifetime for the networks.
4) **Hop Count:** It is defined as the number of hop count is needed to reach a data packet from source node to sink.

### 4.2. Results Discussions

The first performance metric that we consider here is average energy consumption, which is of key importance in battery constrained WSNs. The metric is depicted in Figure 3 for the two environments i.e. static sink and mobile sink. As expected, SiMoCo consumes less energy than static sink. Since in case of SiMoCo, as the sink is mobile to collect the data, the distance between source nodes and sink is minimum. Thus it burns less energy.

Further, among these two cases SiMoCo covers maximum rounds until all the one-hop neighbors of sink died is shown in Figure 4. Similarly, Figure 5 shows rounds until first node die with varying size of network. We have observed that, in the case of static sink, the network lifetime decreases with increasing size of the network. This is due to the fact that in dense network each node, acting as a relay for a higher number of nodes, has to receive and transmit a higher number of packets, which leads to faster energy depletion. In the case of static sink the network lifetime is clearly shorter since the sensor nodes close to the sink always relay the packets of all other nodes, which drains them of their energy quite fast. But in SiMoCo, energy consumption is distributed among the large number of nodes giving increasing network lifetime. It is clear from the figures that SiMoCo results longer lifetime.

The fourth performance metric that we assess is hop count. Here large number of events have been generated randomly throughout the network and number of hop count is needed to reach different sinks in the network. The simulation is carried out for 60 minutes. The result is depicted in Figure 6. It is clear from the figure that for static sink larger hop count value is needed more than mobile sink.

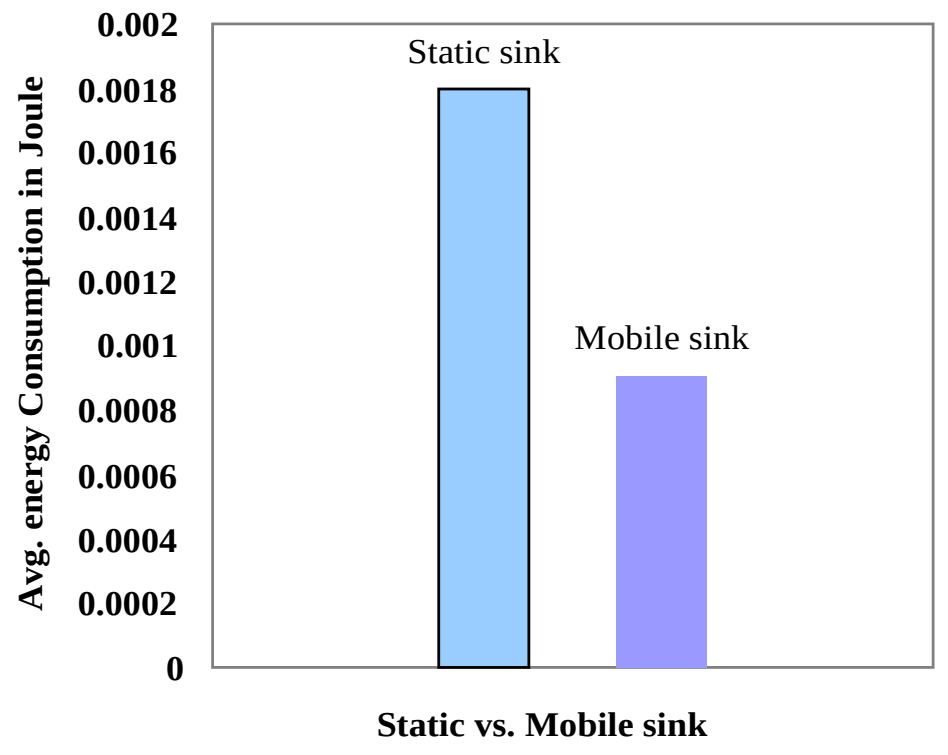

Figure 3. Avg. Energy consumption for static and mobile sink

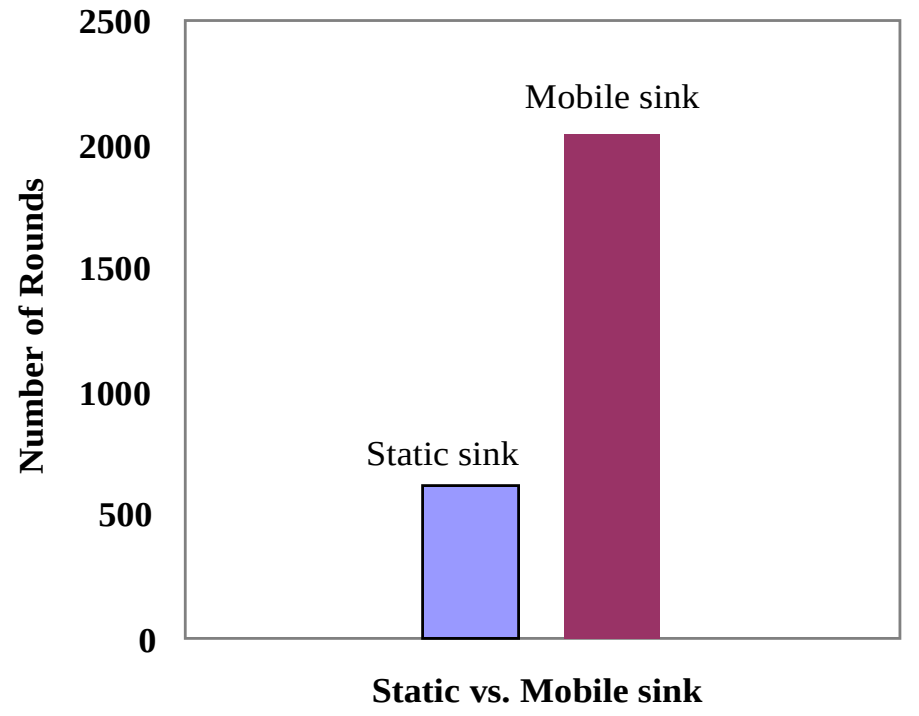

Figure 4. Number of rounds are taken to die all neighbors of the sink

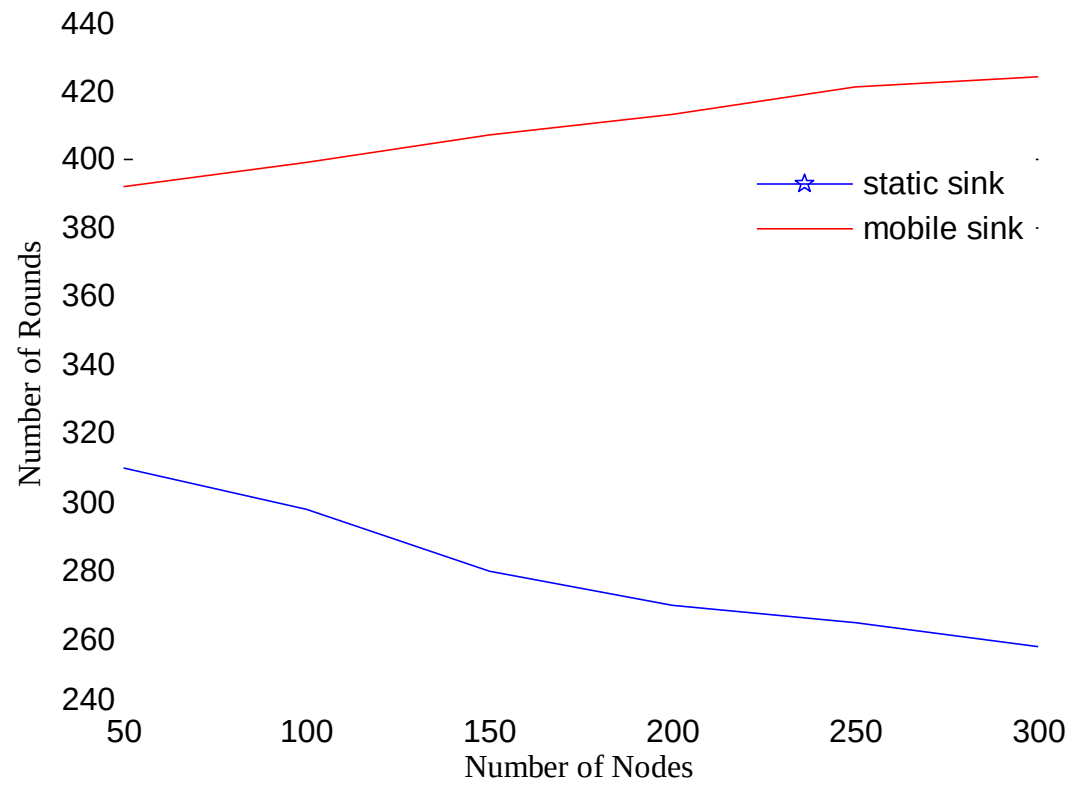

Figure 5. Number of rounds are taken to die first node

The obvious reason is that the communication distance is reduced in mobile sink.

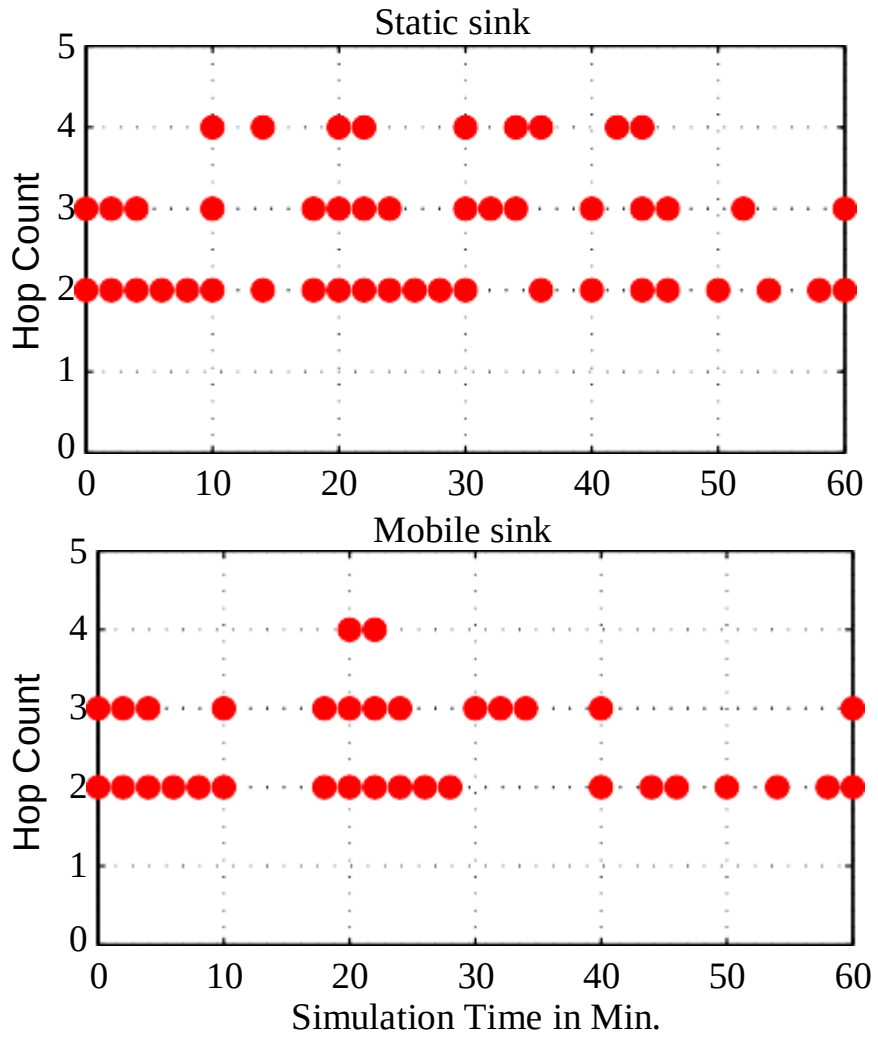

Figure 6. Number of hop count is needed to reach sink for a data packet

Therefore, from the above results, we conclude that by exploiting a mobile sink, network energy is more balanced among the network nodes and data flow bottlenecks at sink can be more effectively avoided, thus improving the overall network lifetime over the static sink environment.

## 5. Conclusion

In this paper a novel sink mobility algorithm with network coverage for routing data packets from a static source to mobile sinks in partitioned multiple sinks wireless sensor networks has been presented. It has been demonstrated that improvement in network lifetime can be achieved by deploying a sink that moves according to the patterns as stated in the proposed algorithm. Energy consumption during data communication is reduced by shortening the source to sink distance. Moreover, Funnelling effect i.e. data flow bottlenecks at sink can be more effectively avoided and ensure network coverage during mobility. From the results it can also be concluded that the SiMoCo shows better performance in all respect when compared with static sink in WSN.